\documentclass[a4paper,11pt]{article}
\usepackage{jcappub}
\usepackage{amsthm}
\usepackage{amssymb}
\usepackage{graphicx} 
\usepackage{dcolumn}
\usepackage{bm}
\usepackage{natbib}
\pdfoutput=1
\usepackage[T1]{fontenc}

\title{\boldmath Reconstruction of cosmic ray air shower core location at SURA experiment\\  }

\author{F.Latifian}

\author{and G.Rastegarzadeh}

\affiliation{Faculty of Physics, Semnan University, Semnan, Iran.}

\emailAdd{f.latifian@semnan.ac.ir}
\emailAdd{grastegar@semnan.ac.ir}

\date{\today}
	
\abstract{SURA is a self-triggered radio array on the roof of physics faculty at Semnan university in Iran. It is designed to detect radio emissions from air showers produced by ultra-high energy (UHE) cosmic rays with energies exceeding $10^{17}$ eV. The array consists of 4 LPDA radio antennas operating in the 40 MHz to 80 MHz range. In this study, we present a method that compares the signal intensities of simulated and experimental data. Specifically, we use a simulated dense array with a large number of antennas as a reference. By comparing the experimental signal intensity of each antenna to that of the corresponding antenna in the reference array, we can reconstruct the cosmic ray air shower core location. We first validate our method on simulated events to estimate the associated error. Afterward, we apply the technique to the cosmic ray candidates detected by the SURA array. Our results show that the core location can be reconstructed with a minimum error of about 6 m. However, when the characteristics of the shower being reconstructed differ significantly from the reference array, the error increases. Finally, we propose optimizations to improve reconstruction accuracy and reduce simulation time.}

\begin{document}

    \maketitle
	\flushbottom

\section{Introduction}
\label{sec:1}

High-energy cosmic rays passing through the atmosphere of the Earth induce showers of secondary particles called extensive air showers (EAS). Despite many efforts done in this field, the origin of ultra-high energy cosmic rays have been an open question.The flux of cosmic rays drop sharply as the primary energy increases. Therefore, to study cosmic rays with energies higher than $10^{15}$ eV, indirect methods, which are based on the detection of secondary particles generated from EAS and analyzing the Lateral Distribution Function (LDF) of showers, should be used \cite{1,2}. To comprehensively investigate this phenomenon, researchers employed various methods for determining the key parameters of cosmic rays including mass composition, primary energy, core location and arrival direction. While the arrival direction of cosmic rays can be measured by time information of secondary particles, determining the mass composition and primary energy are still a challenging task. Different approaches have been used for determining the mass composition and primary energy of cosmic rays \cite{3,4}. An additional parameter that aids in understanding these properties is the depth of shower maximum ($X_{max}$), which represents the depth in the atmosphere where the electromagnetic component of a cosmic ray induced shower reaches a maximum. This parameter has been determined using different approaches, including fluorescence and surface detectors \cite{5}. One of the effective methods of observing high-energy cosmic rays is detecting radio emissions emitted from EAS. In the last decade, various experiments such as LOFAR and AERA have shown that studying some of the EAS key features,  including mass composition, core location, arrival direction and primary energy is possible from radio measurements \cite{6,7}. Following the advances made in radio techniques, the Semnan University Radio Array (SURA) which is located on the roof of Physics Faculty of Semnan University at Semnan city in Iran, started its investigation on the main characteristics of cosmic rays with 4 Log-Periodic Dipole radio Antennas (LPDA). SURA, a well-calibrated radio antenna, has investigated the frequency response of the system and the electric chain  to achieve the best reconstruction of radio signals \cite{8}. Considering the background noise at SURA location, SURA can detect cosmic rays with frequency range from 40 MHZ to 80 MHZ, and the primary energy above $10^{17}$ eV. So far, SURA has successfully determined the arrival direction of cosmic rays, which is one of the important parameters of cosmic rays, by using time information with an accuracy of about one degree \cite{9}. Another main parameter of cosmic rays is the core location  where the shower axis strikes on the earth. Determining the core location is vitally important because it can contribute to determination of other parameters of cosmic rays such as the primary energy and the Lateral distribution function (LDF) of showers. Different methods have used to find the core of showers by various experiments around the world. Both Weighted Average Method(WAM) \cite{10} and the Center of Gravity Method (CGM) \cite{11} were performed by GRAPES-3 experiment in India which have use the secondary particles for measuring the core location . Furthermore, employing small CMOS sensors positioned at specific distances from each other, and calculating the number of charged particles received by each sensor, enables the determination of the core location \cite{12}. In this paper, we reconstruct the shower core at SURA location, by comparing the intensity of the experimental radio signal with the simulated signal in a reference array  and by developing the idea  suggested by CODALEMA experiment \cite{13}. Then, this method is optimized to enhance the reconstruction accuracy and minimize  the simulation time. The contents of this paper are classified into different sections as follows: In section \ref{sec:2}, The design of SURA is described. In section \ref{sec:3}, our method is  introduced.  Section \ref{sec:4} details the simulations performed by CoREAS code. In section \ref{sec:5}, we applied our method to some simulated showers to reconstruct  their core location. Section \ref{sec:6} investigates the effect of the primary energy on finding the core location. Section \ref{sec:7} focuses on optimizing our method using a smaller dense array. In section \ref{sec:8}, we explore the influence of arrival direction of showers on the core reconstruction. Then, In Section \ref{sec:9}, we apply our method to events detected by SURA. Finally, the main results are summarized in Section \ref{sec:10}.

\section{SURA Experiment}
\label{sec:2}

\begin{figure}[!ht]
	\centering
	\includegraphics[width=0.7\linewidth]{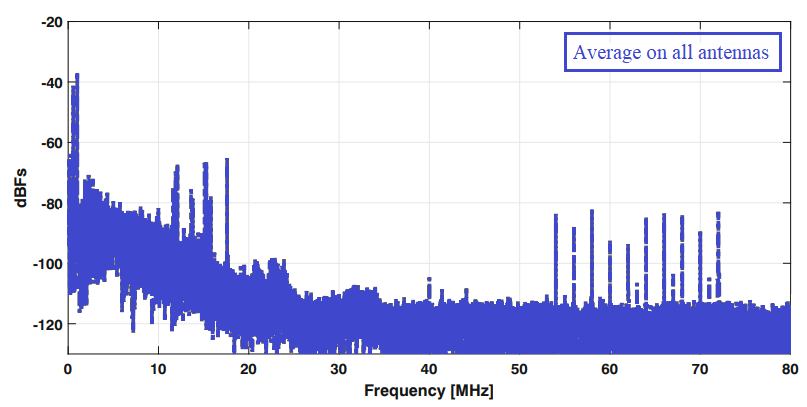}
	\caption{The raw frequency domain of the full spectrum from 0 to 80 MHz at the
		SURA experiment location.}
	\label{fig:1}
\end{figure}

\begin{figure*}[!ht]
	\centering
	\includegraphics[width=0.75\linewidth]{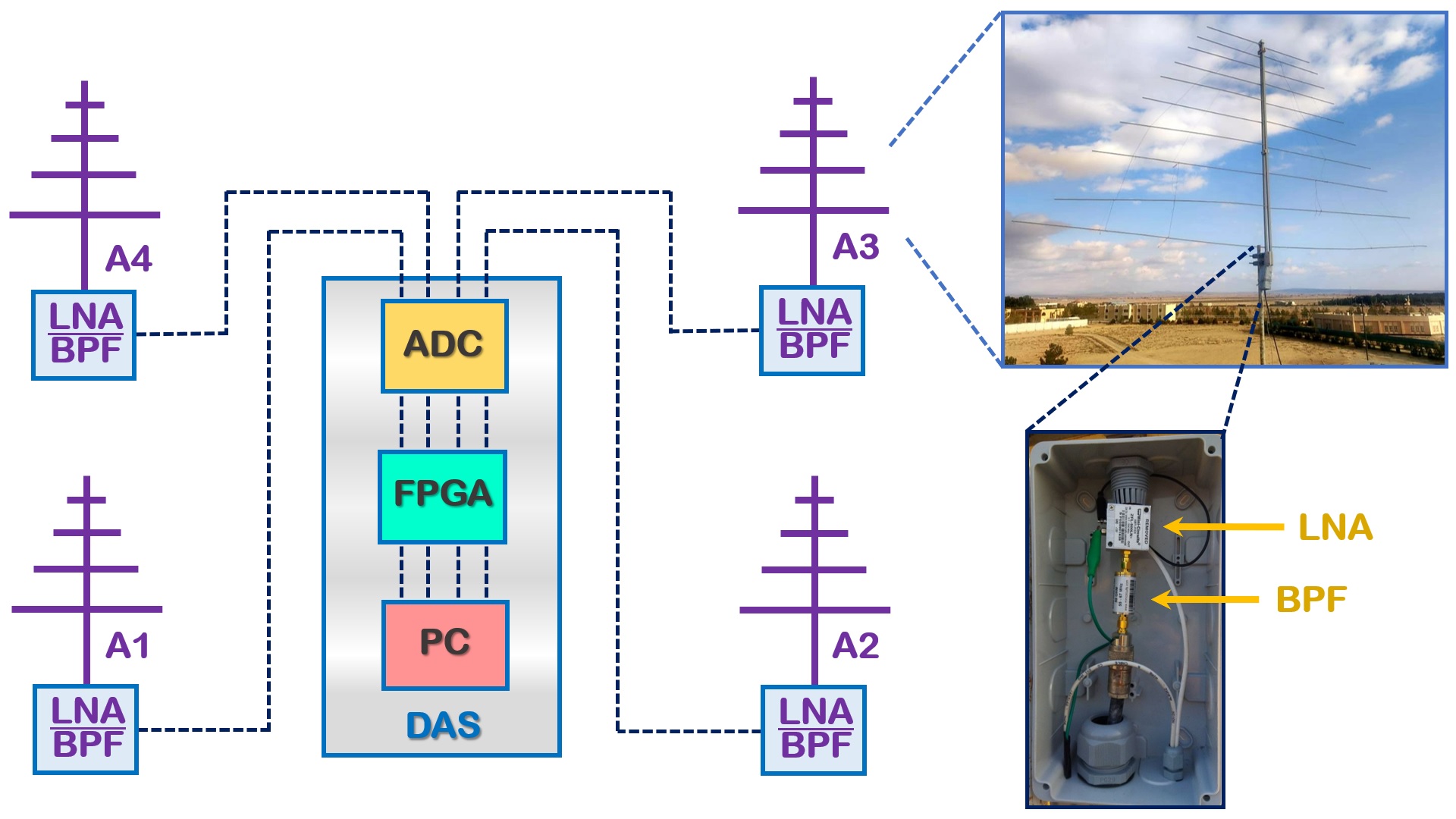}
	\caption{The schematic view of the SURA experiment setup and electronics.}
	\label{fig:2}
\end{figure*}

Semnan University Radio Array (SURA) includes 4 log-periodic radio antennas located on the roof of the Physics Faculty at Semnan University, Iran.
Each antenna consists of ten dipoles made from anodized aluminum alloy, selected for their ultra-corrosion resistance. These dipoles are connected to a central waveguide, positioned on top of each antenna. The dimensions of each antenna are 2.8 $\times$ 5.1 m, and each unit weighs approximately 12 kg. The complete radio spectrum from 0 to 80 MHz at the SURA location is illustrated in figure~\ref{fig:1}. As shown in figure~\ref{fig:1}, the spectrum is highly noisy, particularly in the frequency ranges below 40 MHz. Consequently, we have selected the frequency band from 40 to 80 MHz for SURA operation. Within this operational band, there are a few mono-frequency noises, likely due to TV and radio station activities that are fixed in frequency and do not change over time. To address this issue, we employ notch filters to effectively remove these persistent interferences.  Through computer simulations and by considering the current level of background noise at the experiment's location, we concluded that SURA is capable of detecting cosmic ray candidates with energies above $10^{17}$ eV.  When SURA's antenna receives radio signals, they are initially directed to a low-noise amplifier (LNA) to enhance signal amplitudes. The amplified signals then pass through a Band Pass Filter (BPF) operating within the 40 to 80 MHz frequency range, designed to eliminate unwanted radio emissions.  Both the LNA and BPF are enclosed in a waterproof box mounted under each radio antenna. The amplified and bandpass-filtered radio signals are then transmitted from each antenna to a central data acquisition system (DAS) for further analysis via 50-ohm RG213 Coaxial cable and appropriate connectors. The DAS at SURA consists of a 14-bit, 160 MSPS Analog to Digital Converter (ADC), providing a sampling resolution of 6.25 ns. The digitized signals are then transferred to a Field Programmable Gate Array (FPGA), where an algorithm is used to provide a self-triggering mechanism for detecting cosmic rays at SURA. The algorithm consists of several stages; however the most important ones are discussed here, including a digital high-pass and a series of notch filters to remove unwanted mono-frequency emissions observed in the operative band of SURA. A threshold is also defined, and if the maximum amplitude of a radio signal effectively exceeds this threshold, the event is recorded. Additionally, a specific time window is defined by the algorithm, and only coincident events detected by SURA's antennas within this time frame are stored. If the radio signals meet all the defined criteria, they are saved on a PC connected to the FPGA for further analysis. Figure \ref{fig:2}, shows the schematic view of the SURA setup andb its electronics. We also developed a computer program for offline data processing to select the most probable cosmic ray candidates. For further details on the SURA setup and data processing, please refer to \cite{8,9}.

\section{Core location reconstruction method}
\label{sec:3}

\begin{figure}[!h]
	\centering
	\includegraphics[width=0.7\linewidth,height=0.3\columnwidth]{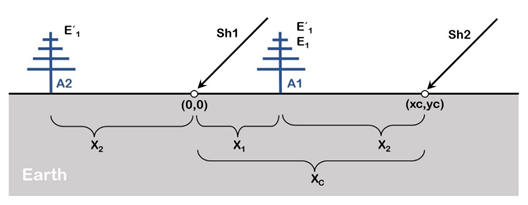}
	\caption{Virtual antennas.}
	\label{fig:3}
\end{figure}

To reconstruct the core location $(x_{c},y_{c})$ of cosmic ray air showers , we compare the intensity of the experimentally observed electric field with the corresponding  electric field in a simulated dense array. The process begins by creating a virtual antenna through simulations, as illustrated in figure~\ref{fig:3}. First, consider an air shower (A1) striking the center of an array at the core location (x=0,y=0) with a specific entry direction. In this case, Antenna 1 detects a radio signal $E_{1}$, from this air shower. Subsequently, if another air shower strikes the array at the core location of $(x_{c},y_{c})$ with the same entry direction, primary mass and energy as the previous one, antenna 1 will receive a different signal, denoted as $E_{1}'$. On the other hand, antenna 2 (A2), which is located at coordinates $x_{2}=x_{1}-x_{c}$ and $y_{2}=y_{1}-y_{c}$, receives the signal intensity $E_{1}'$ from the first air shower with the core location at (x=0,y=0), which is equal to the signal intensity that antenna 1 receives from the air shower with core location at (x=0,y=0). Therefore, we call antenna 2 the "virtual antenna" for antenna number 1. In other words, the signals received from air showers with core location of $(x_{c},y_{c})$ are equivalent to those received from air showers at (x=0,y=0) by their corresponding virtual antenna. Using this principle, we can reconstruct the core location of cosmic ray air showers. To achieve this, a simulation is run for a shower at the core location of (x=0,y=0) impacting a large dense array. We then, compare the signals received from the virtual antennas with the signals obtained from experimental observations, the shower’s core location is determined. To do this, we examine our possible guesses $(x_{i}, y_{j})$ for the core position with one-meter intervals in the range of -30 m $<x_{i}<$30 m and -30 m $<y_{j}<$30 m by using relation \ref{eq:1} and calculating $\chi2$ at each step.

\begin{equation}
	{\LARGE\chi_{ijk} ^2 = \chi^2 (x_{i},y_{j}) =\frac{1}{n} \sum_{k=1}^{n} \left[ \frac{\frac{E^{sim}_{ijk}}{C_{ij}}  - E^{exp}_{k} }{E^{exp}_{k}} \right]^2 }.
	 \label{eq:1}
\end{equation}

where $(x_{i}, y_{j})$ is a core location being tested for the detected events, K is the number of antennas which varies from 1 to 4 for 
SURA, $E^{exp}_{k}$ is the electric field of the antenna number k, $E^{sim}_{ijk}$ is the expected electric field obtained from the simulation for the virtual antenna number k for the possible guesses  of the core location $(x_{i}, y_{j})$ .$C_{ij}$ is a scaling factor which is the mean ratio between simulated signals and experimental data. In this method, when determining the core location, there is no information about the primary energy of extensive air showers. Therefore the intensity of the simulated signal may differ significantly from the intensity of the experimental signal (due to the difference in the primary energy). As a result, they cannot be directly compared with each other. By applying the scaling factor $C_{ij}$, These two signals are adjusted to be comparable. It is clear that if the simulation of the dense array was performed for a shower with lower primary energy than the air shower obtained from the experiment, i.e. $ E^{exp}_{k} > E^{sim}_{ijk}$  , then $C_{ij}$$<$$1$ , otherwise $C_{ij}$$>$$1$. When the value of $\chi^{2}$ is calculated at each step, the tested guess $(x_{j}, y_{j}) $ for the core location that gives the smallest value of $\chi^{2}$ will be the acceptable core location. 

\section{Simulation setup}
\label{sec:4}

\begin{figure}[!ht]
	\centering
	\includegraphics[width=0.6\linewidth]{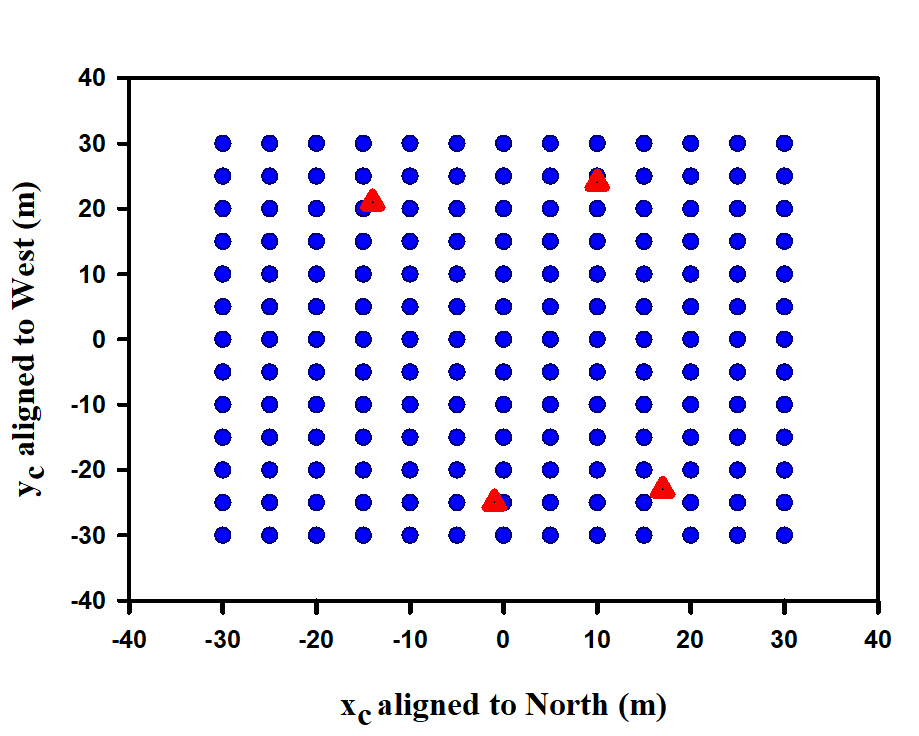}
	\caption{Simulated tested guesses for the core locations
		 ( blue dots) and the SURA radio antennas (red triangles).}
	\label{fig:4}
\end{figure}

Based on the CoREAS 77400 code, we simulate  proton-induced cosmic ray showers while configuring the components of the magnetic field to strengths of $B_{x}$ = 28.09 µT and $B_{z}$ = 39.43 µT, and setting the height above sea level of 11300 meters, corresponding to the values at SURA location. Initially, we simulate a dense array consisting of 12342 radio antennas, which are distributed at a distance of one meter from each other, within a range of -55 m $<$ x $<$  55 m and -55 m $<$ y $<$ 55 m  with a 35$^{\circ}$ zenith angle, a 40$^{\circ}$ azimuth angle, primary energy of $2\times10^{17}$ eV, and core location at ($x_{c}$=0 , $y_{c}$=0). Then, we apply our method to 169 showers with primary energy of $10^{17}$ eV and arrival direction with a 35$^{\circ}$ zenith angle  and a 40$^{\circ}$  azimuth angle. To explore the influence of  primary energy of cosmic rays, we then simulate 507 showers with primary energies of $3\times10^{17}$ eV, $4\times10^{17}$ eV, and $5\times10^{17}$ eV, and an arrival direction with a 35$^{\circ}$ zenith angle  and a 40$^{\circ}$  azimuth angle. Each time, the shower core is varied in 5-meter intervals within the range of -30 m $< x_{i} <$ 30 m and -30 m $< y_{i} <$ 30 m, as shown in figure~\ref{fig:4} (169 showers in each energy). Next, to optimize the method, we simulate a smaller dense array consisting of 3364 radio antennas. This smaller array covers the same area as the initial large dense array, but has antennas spaced further apart compared to the large dense array. In order to use this dense array, 507 simulations are performed with an arrival direction of a $30^{\circ}$ zenith angle and a $18^{\circ}$ azimuth angle, using three primary energies: $ 10^{17}$ eV, $5\times10^{17}$ eV, $7\times10^{17}$ eV. Also, to evaluate the effect of zenith angle on each energy level, we increase the zenith angle by one degree increments from 30° to 35°. Further details about each set of simulations  are provided in the relevant sections.

\newpage

\section{Determining the core location}
\label{sec:5}

\begin{figure*}[!ht]
	\centering
	\includegraphics[width=.33\textwidth,height=5cm]{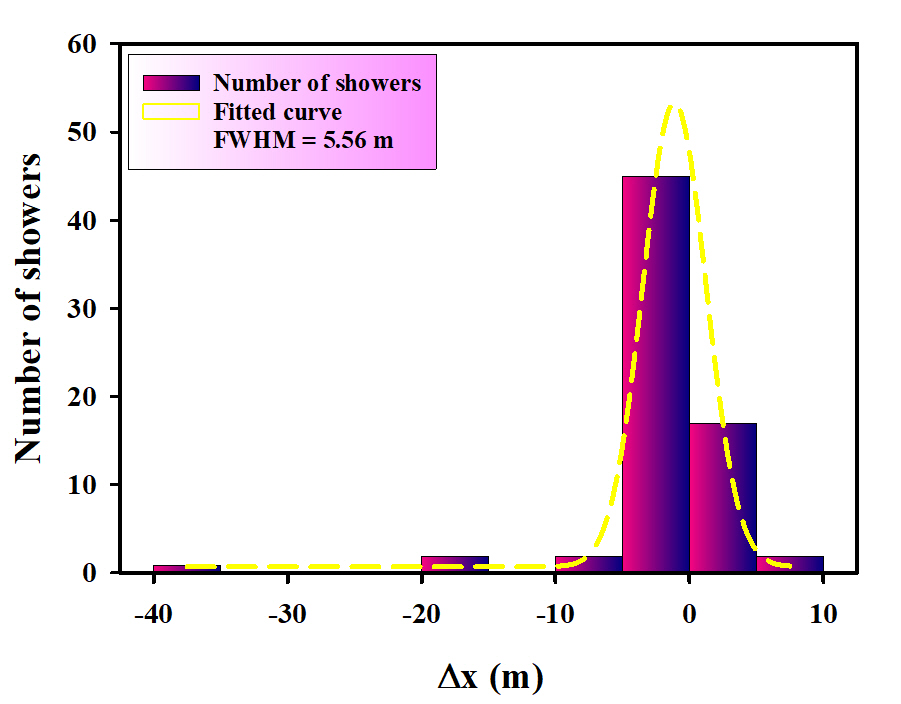}\hfill%
	\includegraphics[width=.33\textwidth,height=5cm]{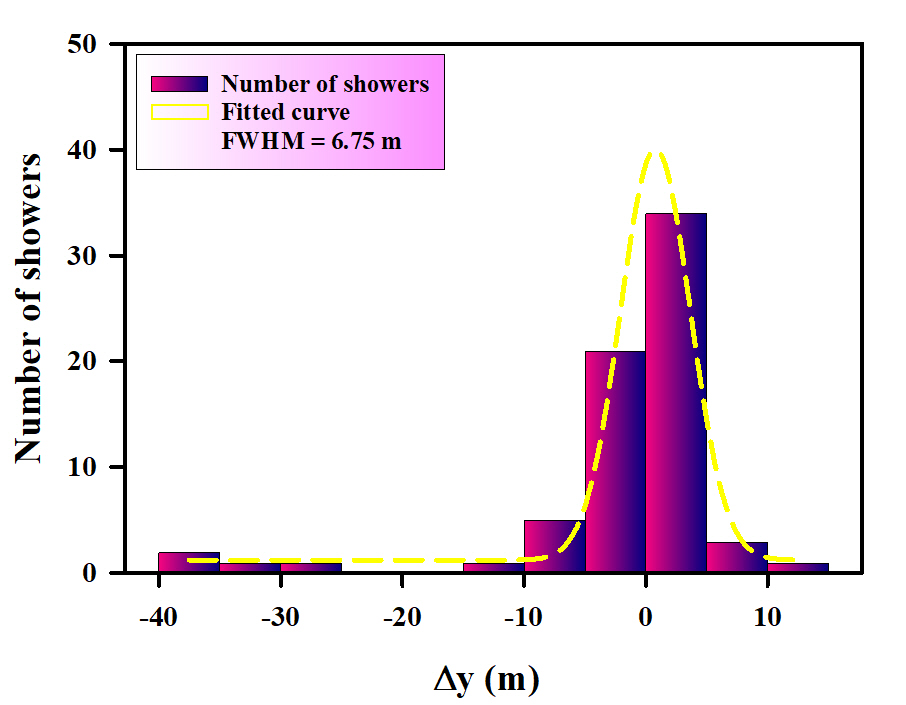}\hfill%
	\includegraphics[width=.33\textwidth,height=5cm]{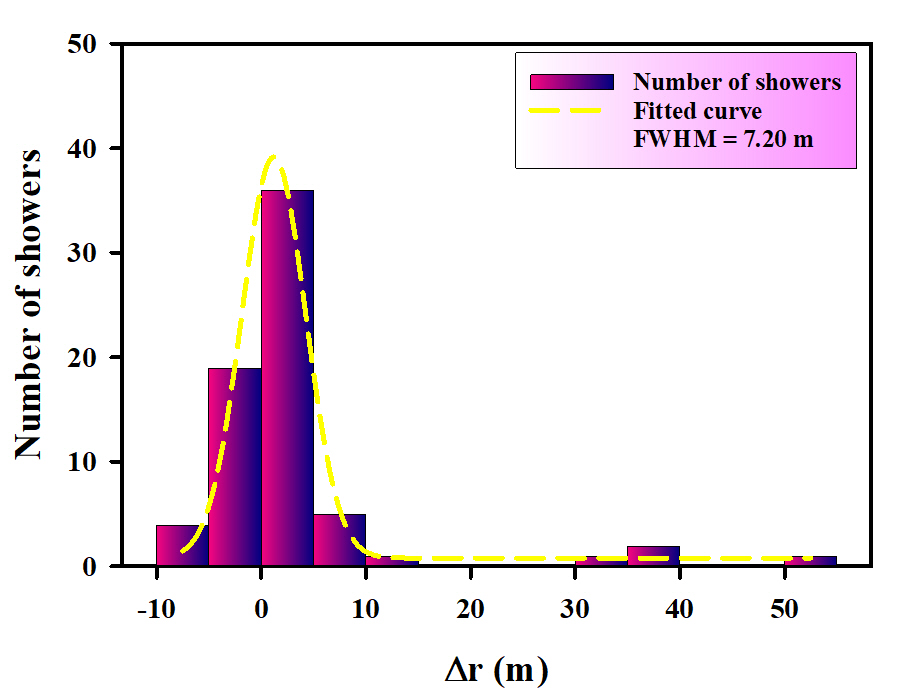}\\
	\caption{Histogram of the error in reconstructing the core location of showers with the primary energy of  $10^{17}eV$, using a dense array with 12321 antennas.}
	\label{fig:5}
\end{figure*}

Using the method described in Section III and employing a large dense array consisting of 12321 antennas, the core location of cosmic ray air showers can be determined by minimizing the $\chi^{2}$ equation. The electric field strength received by the SURA antennas is represented as an $E^{exp}_{ijk}$, while the electric field strength obtained by the antennas in the large dense array is represented as $E^{sim}_{ijk}$ in equation \ref{eq:1}. Through this process, the core of cosmic rays can be reconstructed. It is worth mentioning that in the simulation, the core location is precisely known and entered as an input parameter. However, we assume that the core location is unknown and we attempt to reconstruct it using the proposed method. Subsequently, we compare the reconstructed core location with the known core location from the simulation. By assessing the differences between the reconstructed and actual core locations, we determine the accuracy and reliability of the method. 
Figure \ref{fig:5} illustrates the error in the core reconstruction for one set of simulations, comprising 169 showers with a primary energy of $10^{17} eV$  and an arrival direction with a zenith angle of $\theta=35^{\circ}$ and an azimuth angle of $\phi=40^{\circ}$. Here, $\Delta$x, $\Delta$y and $\Delta$r  represent the error for $x_{c}$ ,$y_{c}$ and r=$\sqrt{x^{2}+y^{2}}$, respectively. These plots can be fitted with the Gaussian function of f(x) = a $exp^{-\frac{(x-x_{0})}{2 \sigma^{2}}}$. In this equation, $\sigma$ is the standard deviation, representing the width of the Gaussian function. From the standard deviation, we can calculate the Full Width at Half Maximum (FWHM) of the Gaussian fitted curve, which represents the errors in the core reconstruction.

\newpage

\section{ Influence of primary energy }
\label{sec:6}
 
\begin{figure*}[!ht]
 	\centering
 	\includegraphics[width=.3\textwidth,height=5cm]{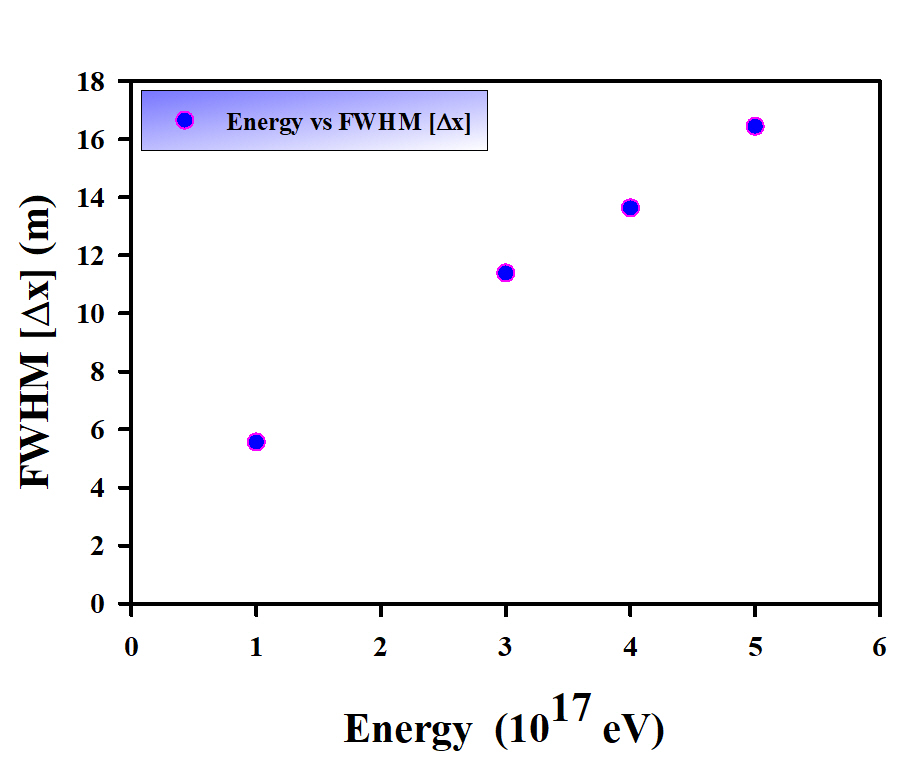}\hfill%
 	\includegraphics[width=.3\textwidth,height=5cm]{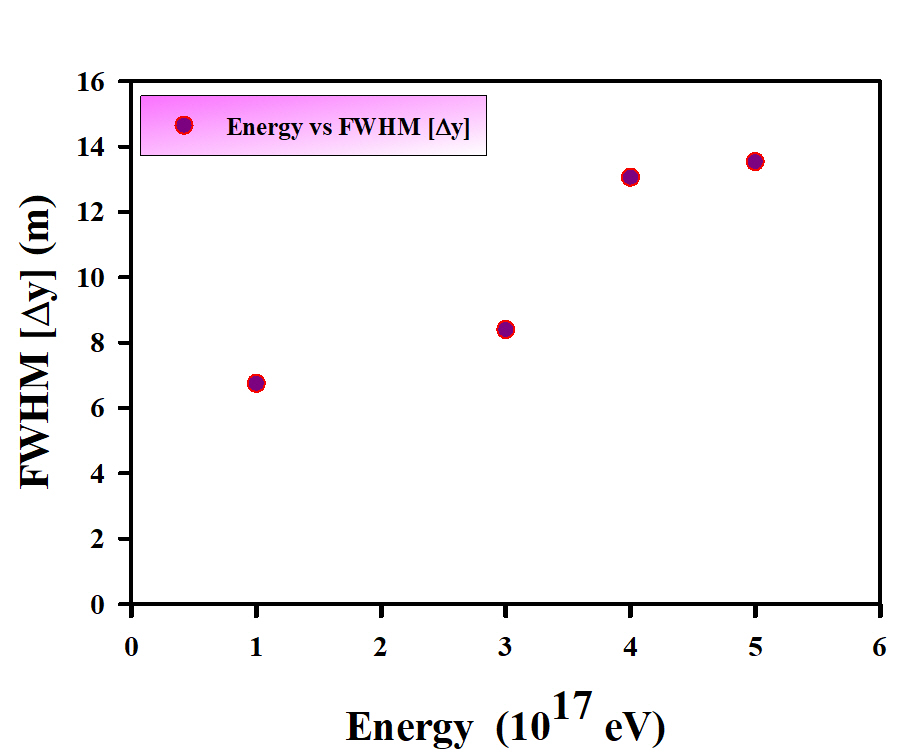}\hfill%
 	\includegraphics[width=.3\textwidth,height=5cm]{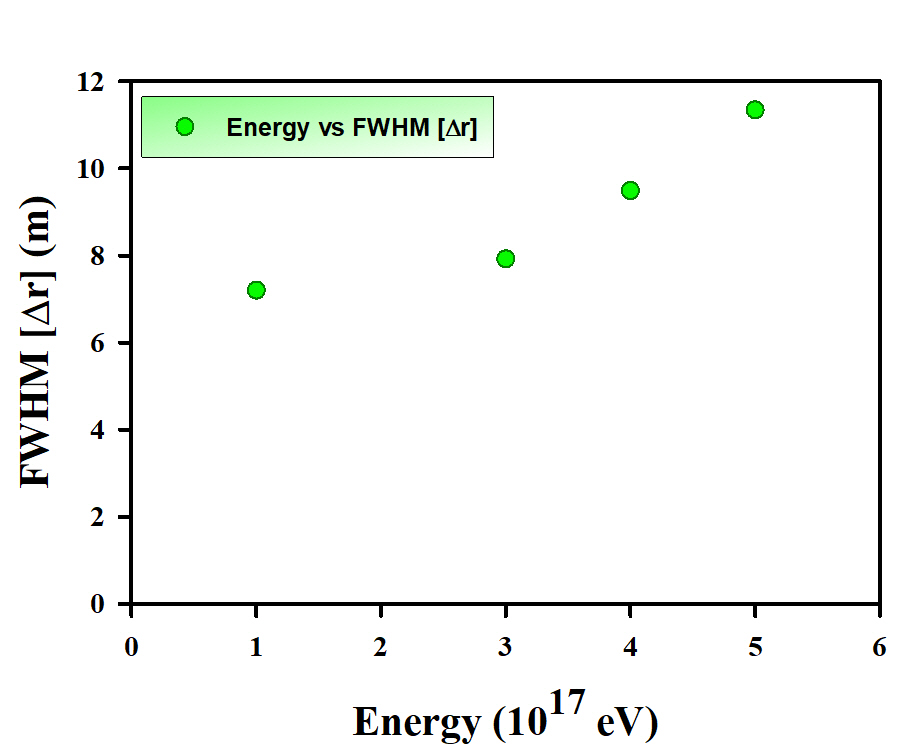}\\
 	\caption{Full Width at Half Maximum (FWHM) of $\Delta x$, $\Delta y$ and  $\Delta r$ distribution versus the primary energy.} 
 	\label{fig:6}
 \end{figure*}

 Obviously, if an air shower strikes SURA with a similar arrival direction, primary particle and energy as the air shower simulated for the dense array, the error in the core reconstruction will be minimal. However, if the primary energy of the shower hitting SURA differs from the shower simulated for the dense array, the error of our method will increase proportionally to their differences. Therefore, we first investigate the influence of primary energy on the reconstruction of the shower core. Thus, air showers with various energies are examined. The errors for reconstructing the core location of showers with different energy levels are summarized in figure~\ref{fig:6}.
  Each data point in the diagram represents the FWHM of the Gaussian function fitted to the histogram of errors for 169 simulated shower cores. In all three diagrams, the error in the calculation of the shower cores increases as the primary energy of showers enhance and deviates from the primary energy of the simulated dense array shower ($2\times 10^{17}$ eV). Consequently, as the primary energy of showers approaches  the primary energy of the simulated dense array shower ($2\times 10^{17}$ eV), the accuracy of determining the core location improves. Figure \ref{fig:6} shows the Full Width at Half Maximum (FWHM) of the errors in the three components, $x_{c}, y_{c}$ and $r_{c}$ of the reconstructed cores locations for showers with energy ranging from $10^{17}$ eV to $5\times 10^{17}$ eV.

\section{Optimization via dense array Size reduction}
\label{sec:7}

\begin{figure*}[!ht]
	\centering
	\includegraphics[width=.33\textwidth,height=5cm]{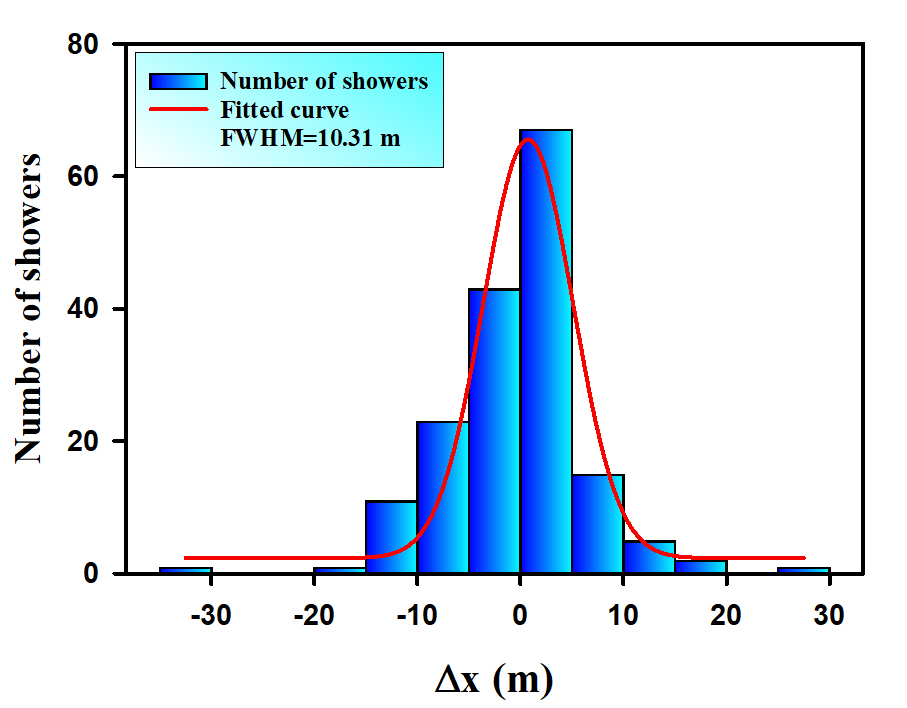}\hfill%
	\includegraphics[width=.33\textwidth,height=5cm]{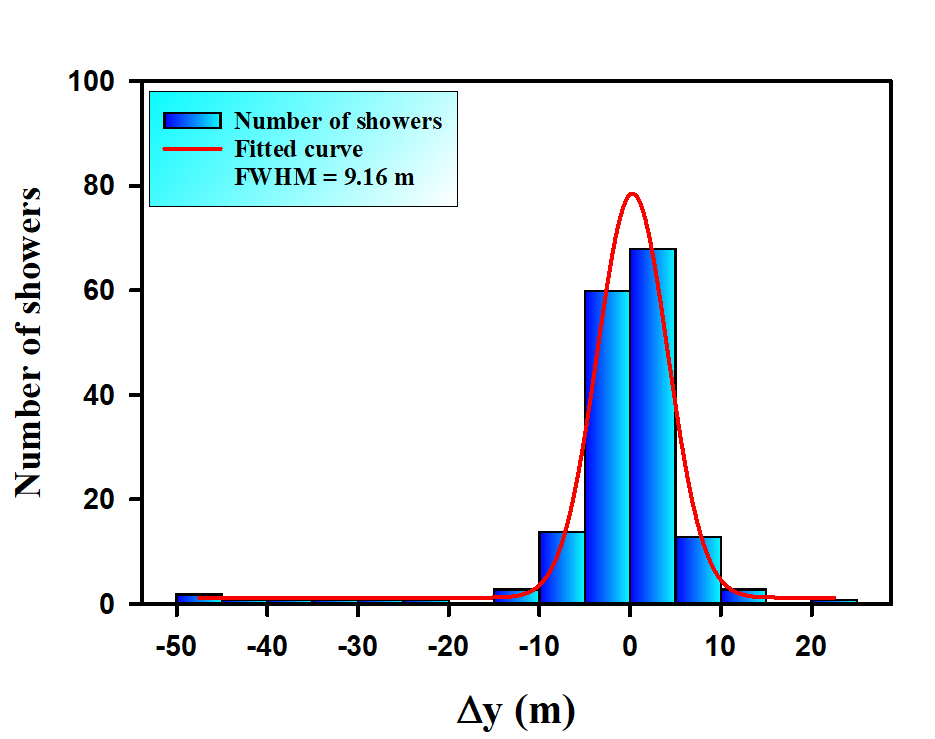}\hfill%
	\includegraphics[width=.33\textwidth,height=5cm]{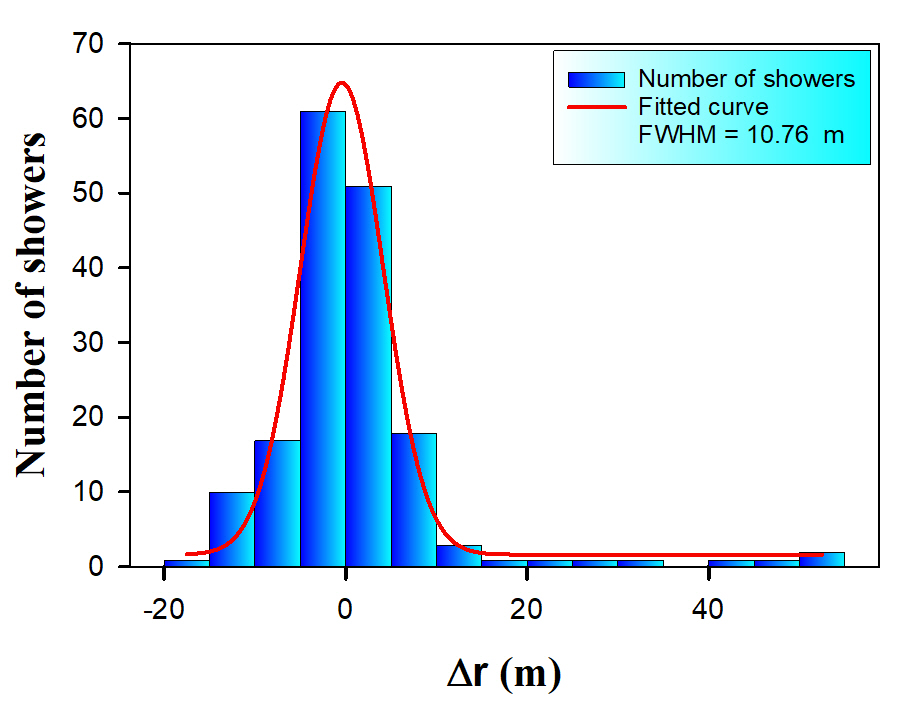}\\
	\caption{Histogram of the error in reconstructing the shower core with primary energy $10^{17}eV$ using a dense array with 3364 antennas.}
	\label{fig:7}
\end{figure*}

One of the challenges associated with the proposed method for determining shower cores is the time-consuming simulation of dense arrays. For instance, simulating a dense array with 12321 antennas requires approximately two months. This method can be optimized by reducing the number of antennas in the dense array. In this section, we simulate a dense array consisting 3364 antennas, which has about  8957 fewer antennas than the dense array used in the previous section. However, it covers the same area, with the aim of decreasing the simulation time and improving our method. We simulate a proton-induced shower with a primary energy of 2$\times10^{17}$ eV, and an arrival direction characterized by a zenith angle of $\theta$ = $30^{\circ}$ and an azimuth angle of $\phi$ = $18^{\circ}$, targeting this reduced dense array. Utilizing this modified dense array, we determine the cores of showers with arrival directions similar to that of simulated dense array, covering energy ranges from $10^{17}$ eV to 7$\times10^{17}$ eV. As depicted in figure~\ref{fig:7}, despite the reduction in the size of the dense array, shower core can be reconstructed with an accuracy comparable to the reconstruction with the larger dense array. The numerical representation of errors at all energy levels is presented in Table \ref{tab:table1}.

\begin{table}[tbp]
	\centering
	\begin{tabular}{|c|c|c|c|}
		\hline
		Primary energy ($10^{17}$ eV) &FWHM [$\Delta$$x_{c}$] (m) &FWHM [$\Delta$$y_{c}]$ (m) &FWHM [$\Delta$$r_{c}$] (m)	\\ 
		\hline
		
		1 & 10.31  & 9.16 & 10.76   \\	
		
		5 & 15.55  & 11.09  & 11.87    \\
		
		7 & 15.18  & 11.36  & 11.26    \\
		\hline
	\end{tabular}
	\caption{\label{tab:table1}FWHM of the Gaussian function fitted on the histogram error of reconstructed shower cores in energy ranging from $10^{17}$ eV to $7\times10^{17}$ eV, using the small dense  array with zenith angle of $\theta=30^{\circ}$ and  azimuth angle of $\phi=18^{\circ}$  }
\end{table}

\section{Influence of arrival direction }
\label{sec:8}

\begin{figure*}[!ht]
	\centering
	\includegraphics[width=.33\textwidth,height=4.2cm]{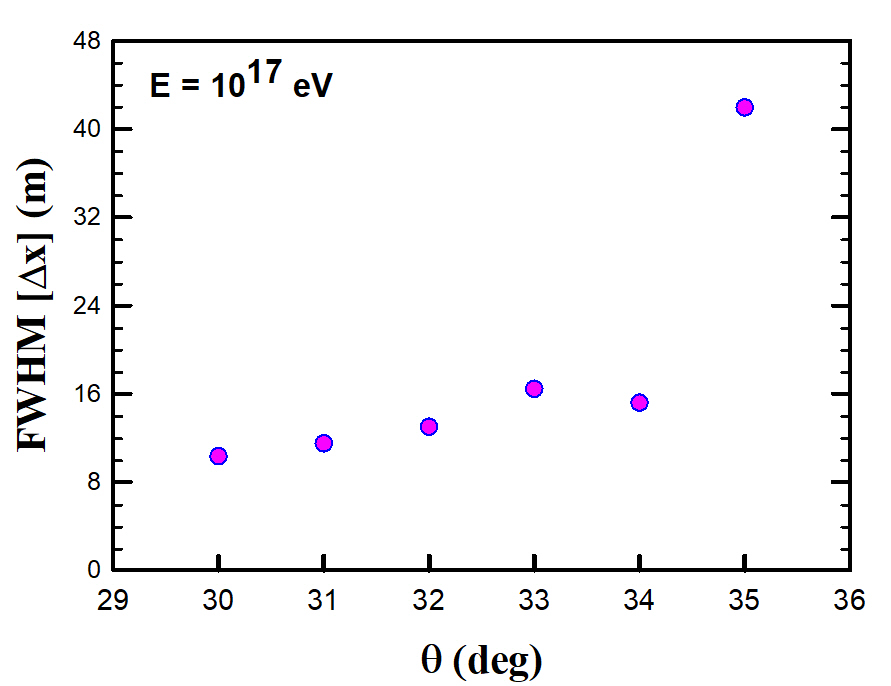}\hfill%
	\includegraphics[width=.33\textwidth,height=4.2cm]{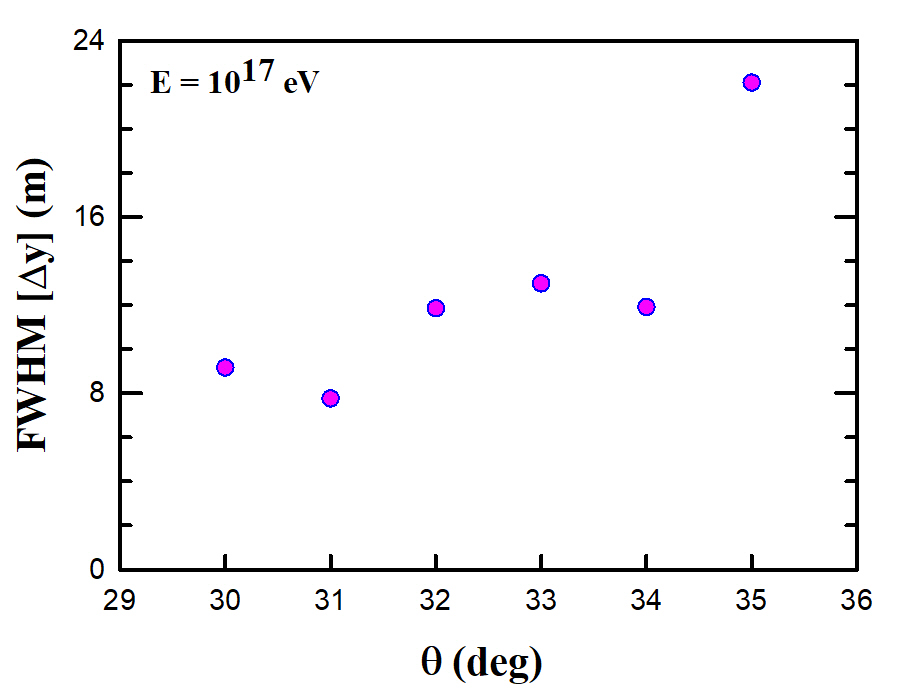}\hfill%
	\includegraphics[width=.33\textwidth,height=4.2cm]{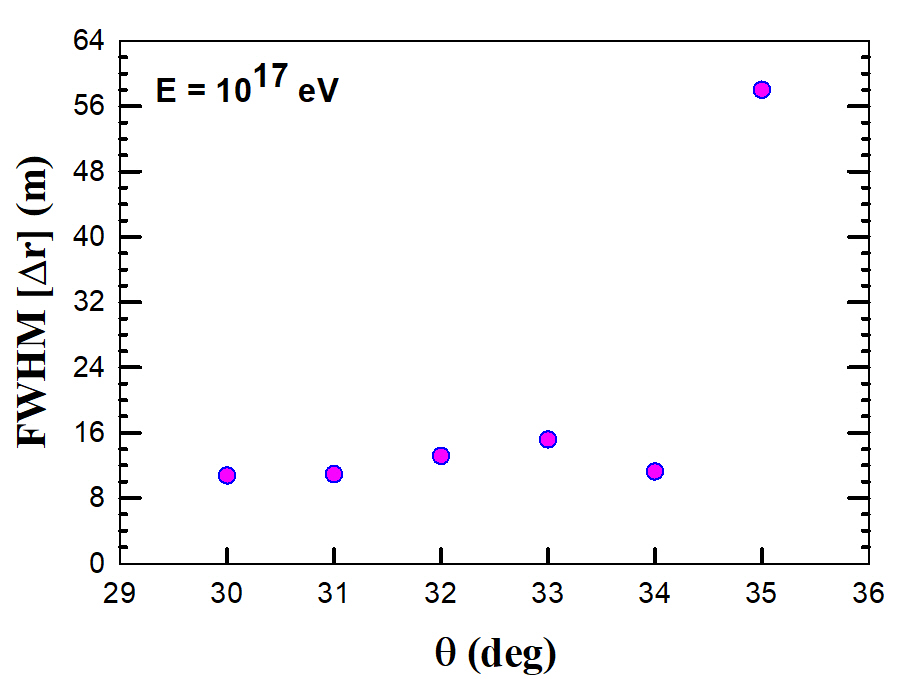}\\
	\includegraphics[width=.33\textwidth,height=4.2cm]{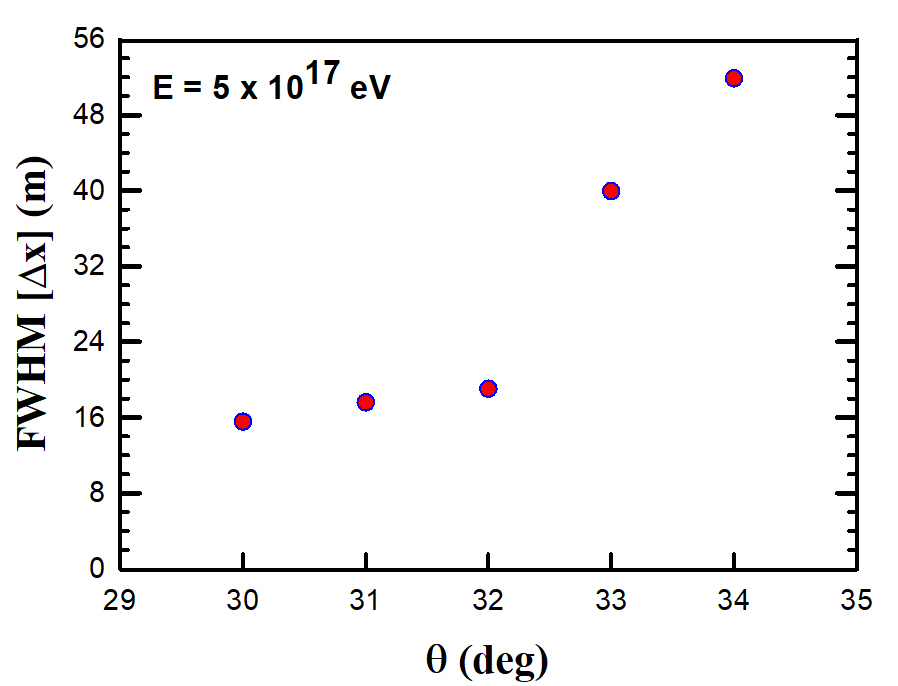}\hfill%
	\includegraphics[width=.33\textwidth,height=4.2cm]{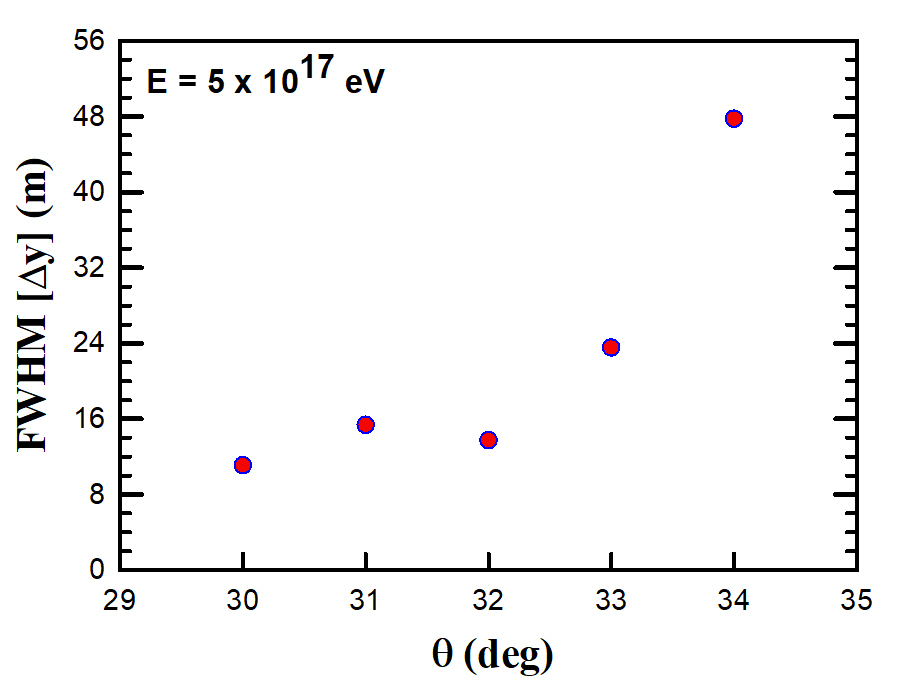}\hfill%
	\includegraphics[width=.33\textwidth,height=4.2cm]{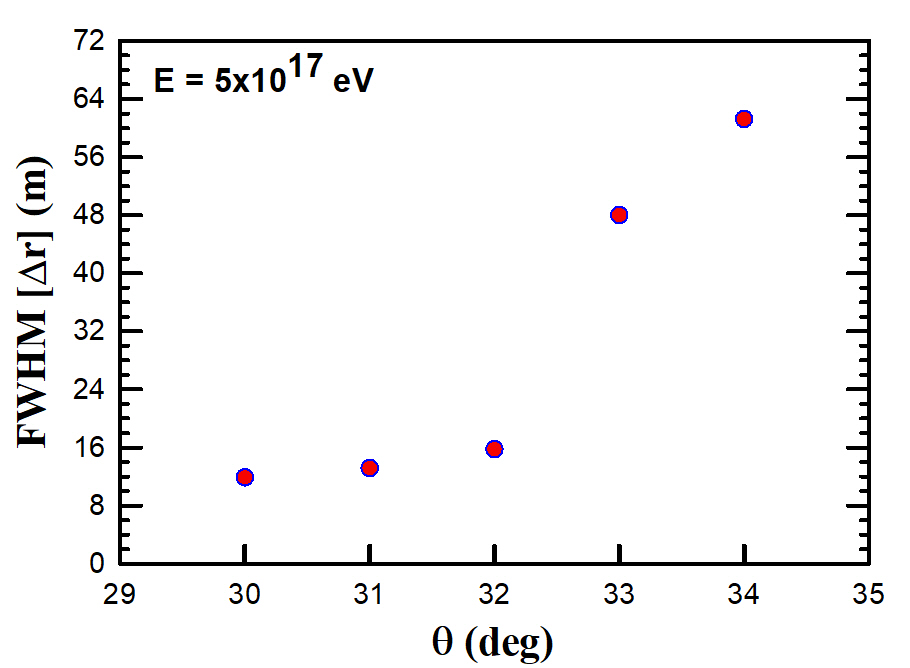}\\
	\includegraphics[width=.33\textwidth,height=4.2cm]{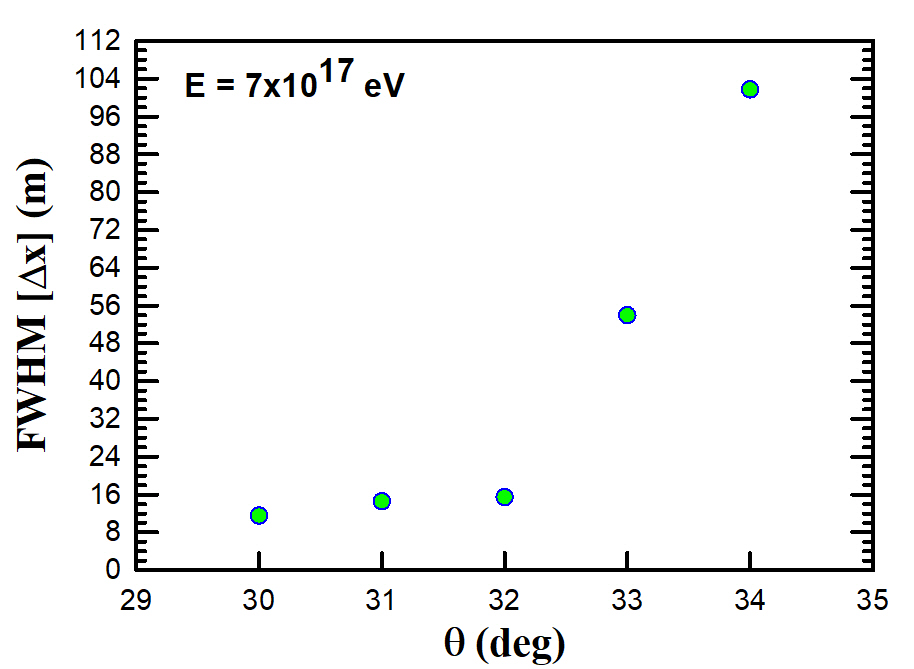}\hfill%
	\includegraphics[width=.33\textwidth,height=4.2cm]{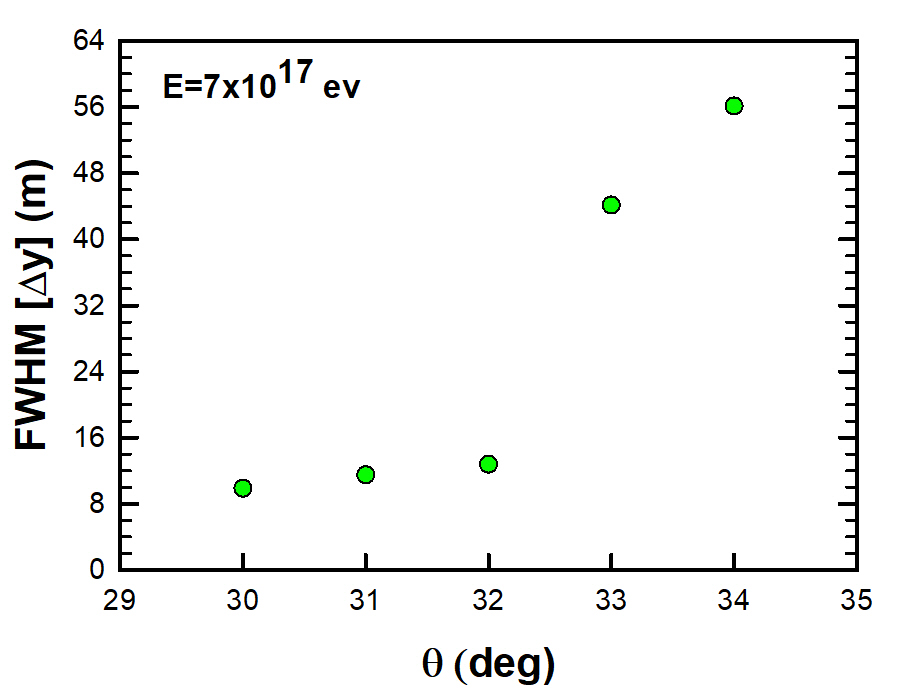}\hfill%
	\includegraphics[width=.33\textwidth,height=4.2cm]{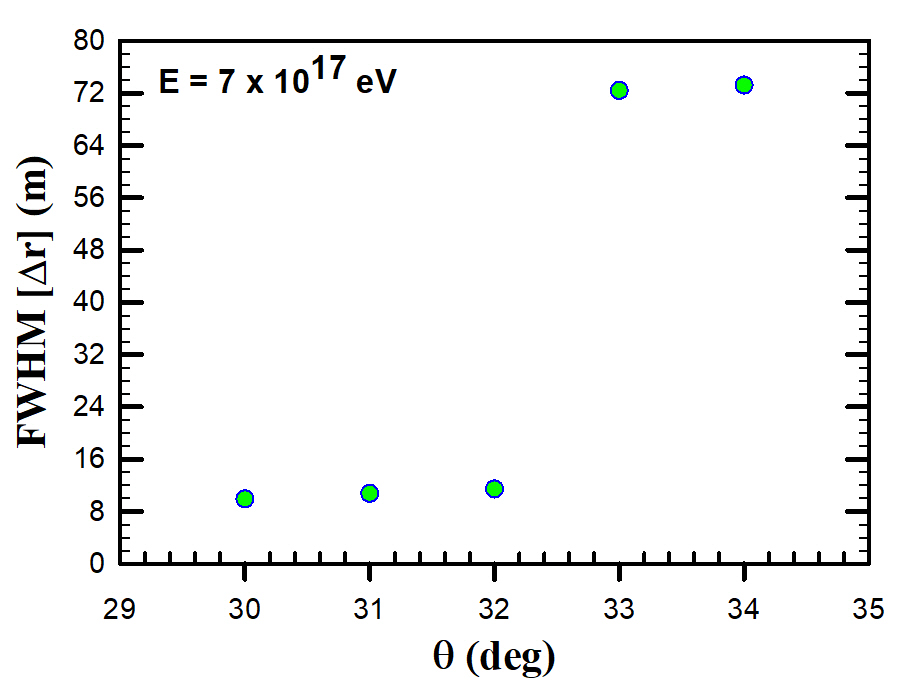}
	\caption{FWHM of $\Delta x$, $\Delta y$ and  $\Delta r$ distribution versus zenith angle ranging from $30^{\circ}$ to $35^{\circ}$.} 
	\label{fig:8}
\end{figure*}

Another limitation of the proposed method is that it can only reconstruct showers with arrival directions aligned with that 
simulated for the dense array.  In other words, reconstructing the core of showers for each arrival direction requires simulating a shower for dense array with a corresponding arrival direction which is a challenging and time-consuming task. To address this issue, air showers with various zenith angles ranging from $30^{\circ}$ to $35^{\circ}$ are investigated at three primary energy levels, including, $ 10^{17}$ eV, $5\times10^{17}$ eV and $7\times10^{17}$ eV. This section explores the impact of arrival direction on core reconstruction accuracy. As shown in figure~\ref{fig:8}, the error increases as the zenith angle rises and deviates from the zenith angle of simulated dense array shower ($30^{\circ}$). Based on figure~\ref{fig:8}, we conclude that for showers with a primary energy of $10^{17}$eV, which is close to the primary energy of the simulated dense array (2$\times$$10^{17}$eV) , extending the zenith angle by up to 4 degrees beyond the zenith angle of the simulated dense array shower ($30^{\circ}$) allows us to use this method and reconstruct the shower core with acceptable accuracy. Therefore, it is unnecessary to simulate dense arrays for showers with these zenith angles. At higher energy levels, this dense array can be used to reconstruct shower with zenith angles extended by only 2 degrees. Overall, this optimization enables us to use a single dense array to determine core locations of showers with various energies and arrival directions.

\section{ SURA Event Selection}
\label{sec:9}

\begin{figure*}[!ht]
	\centering
	\includegraphics[width=0.8\linewidth]{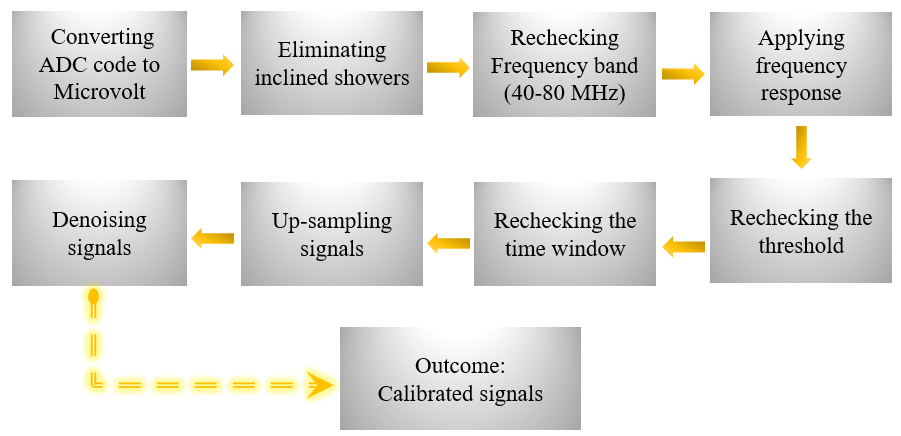}
	\caption{Block diagram outlining the calibration steps of cosmic ray events at SURA experiment.}
	\label{fig:9}
\end{figure*}

\begin{figure}[!htp]
	\centering
	\includegraphics[width=0.55\linewidth]{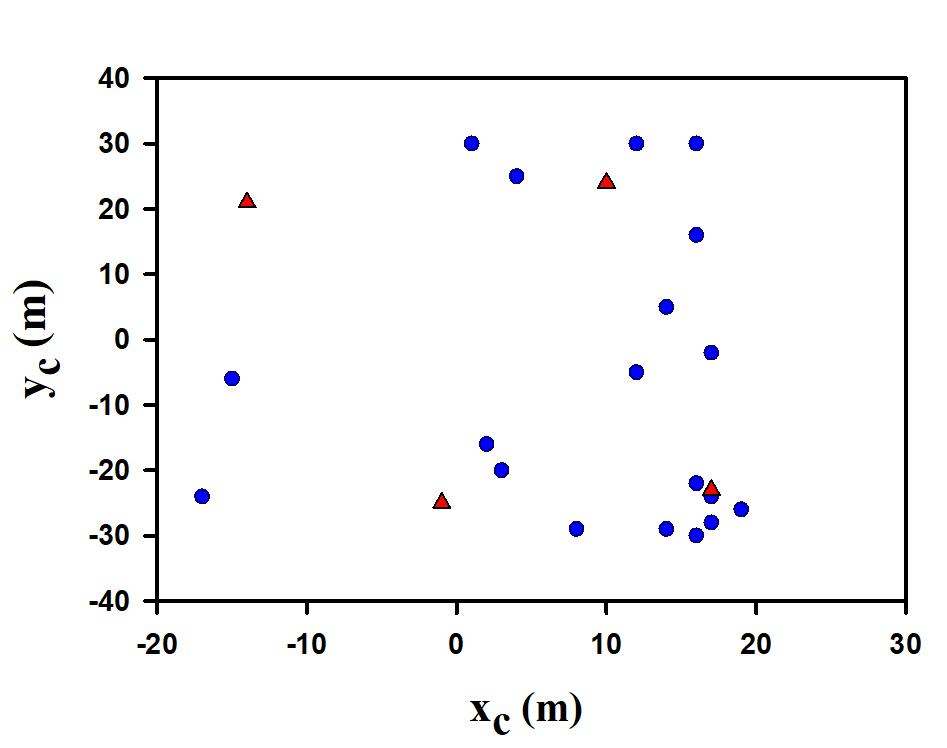}
	\caption{Reconstructed core location of experimental cosmic ray candidates at SURA. red triangles are the SURA's antennas; blue circles are reconstructed cores.}
	\label{fig:10}
\end{figure}

 Radio signals that exceed the background noise threshold at the SURA location and detected by SURA antennas within a specific time window are stored in a dedicated computer system in the cosmic ray laboratory at the Physics Faculty of Semnan University. To ensure high-resolution of signal data, a sampling rate of 160 Mega samples per second is employed. To apply our proposed method for reconstructing the core location of cosmic ray air showers using the experimental data, about 46,646 events captured within a month were carefully selected. Then, the selected events undergo calibration through eight main steps. The calibration process  starts with converting the intensity of radio signals from ADC code to Micro Volt. Additionally, the arrival direction of events is determined by considering the flat wavefront. In step 2, the arrival direction of each event is examined, and inclined events with a zenith angle exceeding $75^{\circ}$ are rejected to ensure we do not include human-made signals. In step 3, a frequency band pass filter is applied to eliminate unwanted radio emissions below 40 MHz and beyond 80 MHz. In the next step, the frequency response of radio antennas and the electronic chain is applied to the radio signals, followed by a recheck of the threshold value and time window. In step 7, the radio signal is up-sampled, converting the original time resolution of 6.25 ns to 0.2 ns using multi-rate signal processing methods to reconstruct the best possible shape of the cosmic ray signals waveform. Finally, the waveform is corrected through a denoising process. The calibration steps applied to the radio signals at the SURA experiment are shown in figure~\ref{fig:9}. After calibration, 136 candidates for cosmic rays remained from the initially selected events, from which, 22 events were chosen to determine the core location. These 22 events were meticulously selected to have arrival directions closely aligned with the large dense array with a zenith angle of 35$^{\circ}$ and an azimuth angle of 40$^{\circ}$. The large dense array was chosen to have a better accuracy in reconstructing the core location of cosmic rays. It is worth mentioning that these selected experimental data are only potential candidates for cosmic rays and may not necessarily be the real cosmic rays. Figure \ref{fig:10}, shows the core location of selected events after calibration.

\section{ Conclusion}
\label{sec:10}

This paper presents a method for determining the core location of cosmic ray air showers by comparing simulated radio signal intensities from a large dense array with the experimental data captured by SURA. The method was tested on different sets of cosmic ray simulations. To optimize this method in terms of simulation time, a smaller dense array with fewer radio antennas, but covering the same area as the large dense array, was employed. It was found that by using this small modified dense array, determining the core location of cosmic rays is achievable with an accuracy comparable to that achieved with the large dense array. Furthermore, to enhance optimization, the influence of the primary energy and the arrival direction was investigated. These investigation revealed that it is unnecessary to simulate multiple dense array for reconstructing the core location of cosmic ray air showers at various energy levels and arrival directions. A dense array can effectively reconstruct the core location of showers with primary energy close to that of the dense array (2$\times$$10^{17}$eV) with  acceptable accuracy. However, as the primary energy of showers deviates from that of the dense array, the error increases. Additionally, we demonstrated that for showers with primary energy close to that of the dense array, extending the zenith angle by up to $4^{\circ}$ beyond the zenith angle of the dense array ($30^{\circ}$) allows us for accurate core location reconstruction. However, further expansion of zenith angle results in  increased errors. On the other hand, as the primary energy of showers deviates from that of the dense array, accurate core reconstruction is only feasible with a zenith angle extension of up to 2 degrees. In our future studies, we aim to minimize the number of antennas in the dense array to the smallest possible number and implement a 3D fitting algorithm for the absent antennas within the dense array to reduce the error and improve our method. In addition, we will analyze other energy levels and arrival directions.

\bibliographystyle{unsrtnat} 
\bibliography{bib}

 \end{document}